# Assessment of Response Time for New Multi Level Feedback Queue Scheduler

M.V. Panduranga Rao[1] , K.C. Shet[2]

[1]*Research Scholar, NITK, Surathkal , Mangalore, India*
[2]*Professor, NITK, Surathkal, Mangalore, India*

*Abstract*— Response time is one of the characteristics of scheduler, happens to be a prominent attribute of any CPU scheduling algorithm. The proposed New Multi Level Feedback Queue [NMLFQ] Scheduler is compared with dynamic, real time, Dependent Activity Scheduling Algorithm (DASA) and Locke's Best Effort Scheduling Algorithm (LBESA). We abbreviated beneficial result of NMLFQ scheduler in comparison with dynamic best effort schedulers with respect to response time.

## I. INTRODUCTION

In a multiprocessing environment of computer, the operating system's theme is to assign the execution time for various processes. The main component of kernel that creates the "selection" is called the scheduler [1][4][31][44][51]. The representation of the policy of scheduler employed is called the scheduling algorithm. Numerous different scheduling algorithms have been deduced over the years.

## II. REVIEW OF DYNAMIC BEST EFFORT REAL-TIME SCHEDULING ALGORITHMS

At least two flavors of best-effort real-time scheduling algorithms exists. This comprises of the Dependent Activity Scheduling Algorithm (DASA) and Locke's Best Effort Scheduling Algorithm (LBESA) [2][6][38][42][49].

In fact, the DASA and LBESA scheduling algorithms are comparable with the canonic Earliest Deadline First (EDF) algorithm during under-loaded conditions. In this case, the EDF promises to meet entire deadlines and is always optimal. However, in the case of an over-loaded circumstance, DASA and LBESA attempt to increase the total task benefit [12][17][34].

The author R.K. Clark et al. [43], demonstrates that DASA more often, overtakes LBESA for the duration of over-loaded circumstances. Peng Li and Ravindran [41], confirms that DASA overtakes the Robust Earliest Deadline First (RED) scheduling algorithm [15][19][35].

Burns A et al. [10]. asserts that alternative features of DASA and LBESA are utilized for the development of MK7.3 kernel. R.K. Clark et al. [43], emphasizes that wide-ranging characteristics of DASA and LBESA are exploited for the growth of Alpha real-time operating system [13][16].

According to R.K. Clark et al. [43], DASA and LBESA are the decent, benefit accrual scheduling algorithms. Both of these algorithms are employed to utmost extent for the growth of mission critical systems [25][26].

The characteristic and performance measurement parameters of each of the scheduler are known from previous work of others. The literature survey provided several outstanding noticeable features [9][14][27][32].

After studying policy mechanisms of different available schedulers, a New Multilevel Feedback Queue (NMLFQ) scheduling algorithm is proposed. NMLFQ includes all necessary modules to compete as a real time scheduler applied in embedded system domain [18][22].

The MLFQ principle of operation is used in NMLFQ scheduling algorithm in such a way that the response time is reduced and the functionality of the scheduling is improved [21][23][33]. The maximum number of queues and the quantum burst time for each queue are found using dynamic method. In NMLFQ scheduling, the operating system can modify the number of queues and the quantum of each queue according to the obtainable processes, deadline and as well as urgency consideration of the process (Garcia P. et al. [16]).

In *Earliest-Deadline-First* scheduling (EDF) priority level for every task is neglected. It primarily concentrates on completion of each task. It focuses on selection of task with nearest deadline for execution. Respectable CPU utilization will be gained, only when the chosen task is having precise deadline [3][20][36][42]. The canonic algorithm for real-time operating systems is EDF. It practices dynamic scheduling principle. The ready process from pool of priority queue is to be selected. The chosen process from the queue will be searched, for the process nearest to its deadline. The selected process is then scheduled to begin execution. The main drawback of the algorithm is, as and when the system is overloaded, the set of processes will miss deadlines and the system leads to deadlock.

Fig. 1, demonstrates the problem of missing of tasks in response to a deadline. The X axis corresponds to the values of time in milliseconds. Y axis corresponds to light weight

processes or small threads. In the example of a load sharing of several time restraint jobs, response is expected at accurate time [5][7][39][47][53]. As the fig. 1, illustrates, request is made at 3 milliseconds. The response is expected within deadline period. Here a dispatch latency of 1 millisecond is suffered during the operation. Henceforth the response is sent after the deadline period. The response is sent at 4 milliseconds missing the deadline, in turn results in invalid response. The data object is passing some vital data to another object involving many processes and threads. We expect to have a response to a request within three milliseconds. The requirement will be perfect if the system reacts within two milliseconds. If the response comes after three milliseconds, the job execution crashes. There exists a sleek time of 1 millisecond from 3 to 4 milliseconds. So here the task missing the deadline results in invalid response.

In general a scheduler is written precisely to respect application priorities. It is necessary for real time applications need to be developed with limited dispatch latency [8][10][28][40][45]. The term *dispatch latency* designates the amount of time a system takes to respond to a request for a process to begin operation. The complete application response time comprises of the interrupt response time, the dispatch latency and the application's response time. The system detects that a process with higher priority than the interrupted process is now ready to dispatch and dispatches the process. The time to switch context from a lower priority process to a higher priority process is comprised in the dispatch latency time.

The shorter version of the pseudo-code of NMLFQ scheduler is described in the following section. This includes dynamic priority of arriving processes. Construction of queue is performed by loading with, certain number of processes to carry out small jobs [11][20][29][37].

Depending on the load of jobs for the scheduler, the numbers of processes are dynamically varied in ready queue. We can designate several levels of queues, depending on waiting condition of processes and urgency conditions of tasks. At any instant of time and with any number of processes and jobs to be completed, scheduler must guarantee meeting the deadlines of the hard real-time tasks.

```
PUBLIC CLASS BEGIN : DynamicPriority
   VARIABLE : Vector<Process> readyQueue =new Vector();
   VARIABLE : Hashtable<Integer, Process> pQueue = new Hashtable<Integer, Process>();
   VARIABLE : public Process p=null;
   VARIABLE : public Process removedProcess=null;
   CONSTRUCTOR BEGIN :  DynamicPriority()
        //Initially loading Queue with some number of processes.
       MONITOR BEGIN :
         FOR BEGIN :(int i=1;i<=pQueue.size();i++)
         //Creating process
              p=Runtime.getRuntime().exec("ps -ef");
              insertIntoPQueue(i*10,p);
         FOR END
       MONITOR END
       CATCH BEGIN (Exception ex)
            ex.printStackTrace();
       CATCH END
   CONSTRUCTOR END
   CONSTRUCTOR BEGIN : public DynamicPriority(int i)
      MONITOR BEGIN :
         //Removing the requested number of processes from Queue
         OUTER IF BEGIN (i<10)
         FOR BEGIN (int j=0;j<i;j++)
            removedProcess=removeAtEnd(j);
            INNER IF BEGIN (removedProcess != null)
                System.out.println("process is removed from queue");
            INNER IF END
            INNER ELSE BEGIN
                System.out.println("process removed from the queue is null");
            INNER ELSE END
         FOR END
         OUTER IF END
         OUTER ELSE BEGIN
              addAtFront(i, Runtime.getRuntime().exec("ps -ef"));
         OUTER ELSE END
      MONITOR END
      CATCH BEGIN :(Exception ex)
           ex.printStackTrace();
      CATCH END
   CONSTRUCTOR END

/* This method add the process to the readyQueue based on priority. It will check the priority and moves the other processes in readyQueue accordingly. */
 METHOD BEGIN : public void addAtFront(int priority, Process temp)
          Enumeration enumitr = pQueue.keys();
      WHILE BEGIN :(enumitr.hasMoreElements())
       IF BEGIN : (!(priority <(Integer)enumitr.nextElement()))
            pQueue.put(priority, temp);
            readyQueue.add(temp);
       IF END
      WHILE END
 METHOD END

/* This method will insert some processes initially into Queue and also readyQueue(implementation details)   */
 METHOD BEGIN : public void insertIntoPQueue(int priority, Process process)
          //Adding process to the Queue and also into readyQueue
```

```
    IF BEGIN :(process != null)
        pQueue.put(priority, process);

        readyQueue.addAll(pQueue.values());
        System.out.println("process is added into Ready Queue");
    IF END
    ELSE BEGIN
        System.out.println("process is null");
    ELSE END
METHOD END

/*  This method removes the process from the Queue.  */
  METHOD BEGIN : public Process removeAtEnd(int index)
        //Removing the process from the index specified from both readyQueue and pQueue
        readyQueue.remove(index);
    RETURN : return(pQueue.remove(index));
  METHOD END

  /* Algorithms guarantee that if a task is accepted for execution, the task and all previous tasks accepted by the algorithm will meet their time constraints [22][26][30][48]. The planning based algorithms attempt to improve the response and performance of a system to aperiodic and soft real-time tasks while continuing to guarantee meeting the deadlines of the hard real-time tasks. */

  METHOD BEGIN : public void planningBased()
    IF BEGIN (true)
        //Task is accepted for execution
        //Resource are allocated.
        //Responds in time
    IF END
  METHOD END

  METHOD BEGIN : public void bestEffortBased()
    IF BEGIN : (true)
    MONITOR BEGIN
        Enumeration enumitr = pQueue.keys();
        Vector sorted = new Vector();

        WHILE BEGIN : (enumitr.hasMoreElements())
            int dynArr =(Integer)enumitr.nextElement();
            sorted.add(dynArr);
        WHILE END
        Collections.sort(sorted);
        FOR BEGIN (int i=0;i<sorted.size();i++)
            //Before calling shortest job first set the priority for the job.
            ExecutingProcess((Process)sorted.get(i++));
        FOR END
    MONITOR END
    CATCH BEGIN (Exception ex)
        ex.printStackTrace();
    CATCH END
        IF END
METHOD END

/* Actual execution started.  */
  METHOD BEGIN : public void ExecutingProcess(Process process)
        //Actual Execution happens
        System.out.println("Execution started...");
  METHOD END

/* We are creating pQueue as a placeholder for processes which are ready to execute. As it is DynamicPriorityQueue, we have to use HashTable for storing both priority and processes. Processes will be added based on the priority to the Queue and will removed from the end or again, based on priority form [24][46][50]. */
  METHOD MAIN BEGIN : public static void main(String[] args)
    // TO-DO Auto-generated method stub
    //Creating Queue
    CALLING METHOD : new DynamicPriority();
        //Removing processes from Queue
    CALLING METHOD : new DynamicPriority(2);
    CALLING METHOD : new DynamicPriority(110);
    MONITOR BEGIN
        CALLING METHOD : new DynamicPriority().addAtFront(2,Runtime.getRuntime().exec( "ps -ef"));
        CALLING METHOD : new DynamicPriority().planningBased();
        CALLING METHOD : new DynamicPriority().bestEffortBased();
    MONITOR END
        CATCH BEGIN(IOException e)
          // TO-DO Auto-generated catch block
          e.printStackTrace();
        CATCH END
    METHOD MAIN END
CLASS END
```

A job is a function of many tasks or processes. Division of jobs into task modules is performed from analysis of jobs. All the steps of NMLFQ scheduler are summarized in fig. 2. This model depicts the coordination amongst various modules of scheduler. Jobs are divided into task modules after the analysis of the nature of jobs. The history of jobs is maintained for future use. The processes are allocated to lock the resources with the control of interrupt handler. The task controller is made to govern for time stamping of deadline. It also supervises ready, pending, blocked and sleeping tasks. The evaluation of priority levels of processes, ordering of processes and division into time slices is controlled by scheduler through the task controller. The peripherals are associated to the CPU through input output controller. This occasionally makes use of context switching. Several queues are supervised by a substantial queue controller. As stated earlier, mapping of processes to critical resource is achieved by semaphores. Dynamic prioritization is a vital task of scheduler, which necessitate raising or scaling of priority of processes.

### III. NMLFQ COMPARISON WITH BEST-EFFORT REAL-TIME SCHEDULING ALGORITHMS - DASA AND LBESA

The NMLFQ real time scheduler is compared with existing best-effort real-time scheduling algorithms. This comprise of the Dependent Activity Scheduling Algorithm (DASA) and Locke's Best Effort Scheduling Algorithm (LBESA) [20][52][54]. The comparison of NMLFQ scheduler with DASA and LBESA is accomplished for twenty test cases. The results of NMLFQ, DASA and LBESA for twenty different set of inputs are reckoned. The results provided ameliorate characteristics for CPU utilization, overall turnaround time, average turnaround time, average waiting time and average response time of schedulers. The comparative results are analyzed for several processes correspondingly, as shown in fig. 3 and generalized graph of hundreds of processes can also be drawn. Fig. 3, exemplifies, Average Response time for NMLFQ scheduler, compared with real time DASA and LBESA scheduers for twenty testcases, proves 10 to 25% reduction of response time. In this research paper, We have depicted the performance with respect to Average Response time for NMLFQ scheduler

Fig. 3, proves as per the achieved results, average response time for NMLFQ scheduler, compared with real time DASA and LBESA scheduers for twenty testcases, illustrates 10 to 25% reduction of response time in each subsidiary stage.

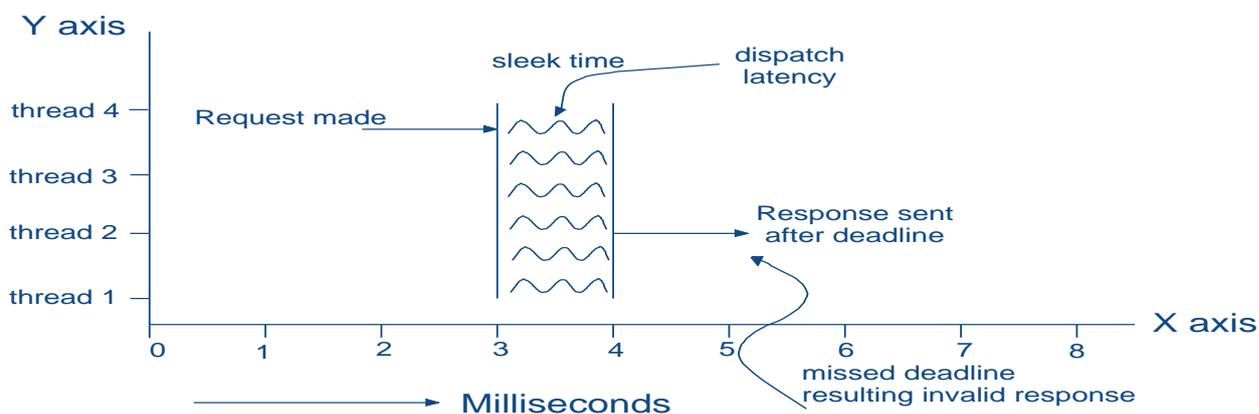

Fig. 1 Task missing the deadline results in invalid response.

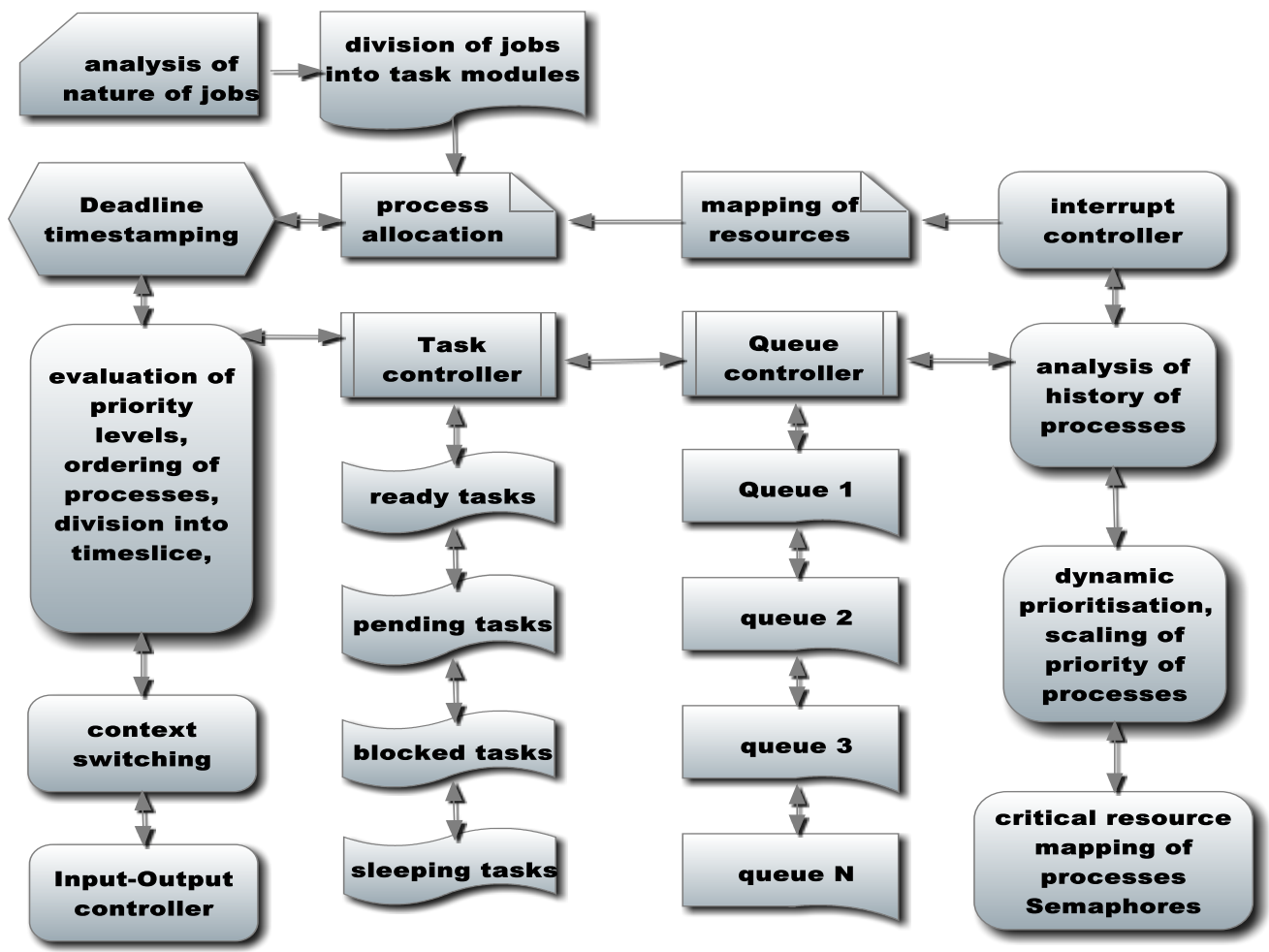

Fig. 2  Model of detailed conceptual view of NMLFQ including several modules.

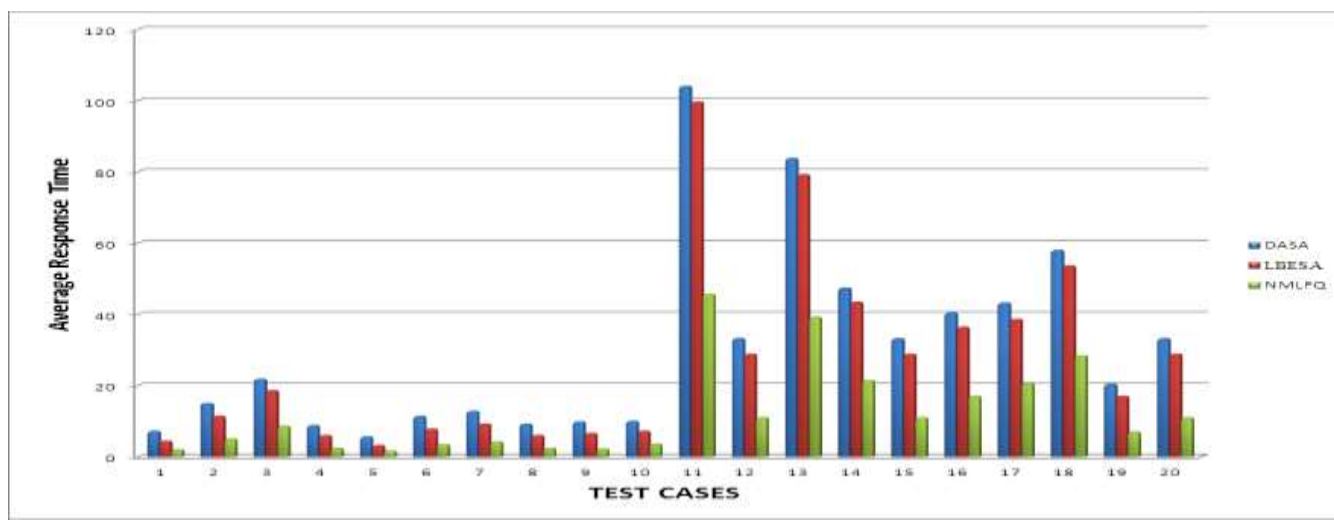

Fig. 3  Average Response time for NMLFQ scheduler, compared with real time DASA and LBESA scheduers for twenty testcases, depicts 10 to 25% reduction of response time in each subsidiary stage.

IV. CONCLUSIONS

In this research paper, we have discussed the themes associated with NMLFQ scheduler. The basic review of dynamic best effort real-time scheduling algorithms is explained. Comparison of proposed scheduler is made, with best-effort real-time scheduling algorithms - DASA and LBESA. Eventually, Model of detailed conceptual view of NMLFQ including several modules is also discussed in short.

V. AUTHOR BIOGRAPHIES

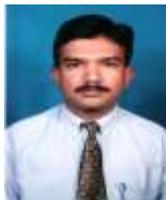

**Prof. M.V. Panduranga Rao** is a research scholar at National Institute of Technology Karnataka, Mangalore, India. He has completed Master of Technology in computer science and Bachelor of Engineering in electronics and communication.

His research interests are in the field of Real-Time and Embedded Systems on Linux platform. He has published various research papers in journal and conferences across India, Also in the IEEE international conference in Okinawa, Japan. He has authored two reference books on Linux Internals. He is the Life member of Indian Society for Technical Education and IAENG.

His webpage can be found via
http://www.pandurangarao.i8.com/ .

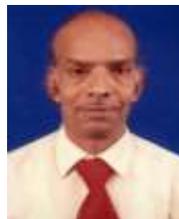

**Dr. K.C.Shet** obtained his PhD degree from Indian Institute of Technology, Bombay, Mumbai, India, in 1989. He has been working as a Professor in the Department of Computer Engineering, National Institute of Technology, Surathkal, Karnataka, India, since 1980.

He has published over 200 papers in the area of Electronics, Communication, & computers. He is a member of Computer Society of India, Mumbai, India, and Indian Society for Technical Education, New Delhi, India.

His webpage can be found via
http://www.nitk.ac.in/~kcshet/index.html .